\documentclass[osajnl,twocolumn,showpacs]{revtex4}
\usepackage{bm}

\usepackage{graphicx,amsmath}
\usepackage{subfigure}

\newcommand{\op}[1]{\pmb{#1}}

\newcommand{\ket}[1]{|{#1}\rangle}

\begin{document}

\normalsize

\title{Single-photon interferometry with orbital angular momentum circumvents standard wave-particle duality}

\author{Michal Kol\'{a}\v{r}$^{1}$, Tom\'a\v s Opatrn\'{y}$^1$, and Gershon Kurizki$^2$}

\affiliation{$^1$Department of Optics, Palack\'{y}
University, 17. listopadu 50,
77200 Olomouc, Czech Republic\\
$^2$Weizmann Institute of Science, 76100
Rehovot, Israel
}

%\pacs{03.65.Ud, 03.65.Vf, 03.75.Dg}
%\pacs{120.3180,  %Interferometry
%050.5080  %Phase shift
%}
\date{\today}
\begin{abstract}
A polarized photon with well-defined orbital angular momentum that emerges from a Mach-Zehnder interferometer
(MZI) is shown to circumvent wave-particle duality. Its polarization-resolved detection probability forms a non-sinusoidal interferometric pattern. 
For certain phase differences between the MZI arms, this pattern yields both reliable
which-path information and high phase-sensitivity.
\end{abstract}

\maketitle
%%%%%%%%%%%%%%%%%%%%%%%%%%%%%%%%%%%%%%%%%%%%%%%%%%%%%%%%%%%%%%%%%%%%
The wave-like behavior of single quantum particles is suppressed whenever we try to learn more
about their complementary particle-like property. For two-path interferometers, e.g.,
the Mach-Zehnder interferometer (MZI), a relation describing such complementary
behavior was derived\cite{Wooters,Vaidman,Englert} and subsequently experimentally tested for
massive particles\cite{Rempe} as well as for photons\cite{Kwiat}. This relation sets an upper bound
on possible \emph{joint} observation of particle-like properties, quantified by the paths' 
{\em distinguishability} \cite{Vaidman,Englert}, and wave-like properties, quantified by the {\em visibility}, 
which is very closely related to the {\em sensitivity} of the fringes to small phase changes. This duality
relation reduces to a tradeoff between our which-way and which-phase knowledge, the latter being our
ability to discern different phase shifts between the MZI arms.
In this Letter, we propose a scheme that is predicted to yield, counterintuitively, considerably more accurate
results for simultaneous path-distinguishability and phase-sensitivity of a photon emerging from a MZI 
than those predicted by the conventional complementarity relation\cite{Vaidman,Englert}.

In refs.~\onlinecite{NJP07}-\onlinecite{PRL06}, we have introduced the concept of 
translational-internal entangled (TIE) states, wherein the entanglement between two non-degenerate orthogonal internal states 
of the particle and two different momentum states, makes their superposition coefficients
dependent on the path length $x$:
\begin{eqnarray}
\nonumber
\ket{\Psi_{\rm TIE}}&=&c_1\ket{k_1}\ket 1+c_2\ket{k_2}\ket 2,\\
\langle x\ket{\Psi_{\rm TIE}}&=&c_1\exp{(ik_1x)}\ket 1+c_2\exp{(ik_2x)}\ket 2,
\label{psi-tie}
\end{eqnarray}
where $c_1$, $c_2$ are the probability amplitudes and $\ket{k_1}$, $\ket{k_2}$ are the wavenumber (momentum) eigenstates.
We assume that the {\em total} energy is the same for both terms in the superposition in (\ref{psi-tie}). 
This ensures that no time-dependent oscillations of these terms take place.

If a particle in TIE state (\ref{psi-tie}) traverses the MZI, its detection probability in one of the MZI output ports,
e.g., the one labeled by ($+$), which receives the sum of the two arms' contributions, can be calculated to be \cite{NJP07,IJPB06}
\begin{equation}
P_+=\frac{1}{2}+\frac{|c_1|^2\cos{(k_1L)}+|c_2|^2\cos{(k_2L)}}{2},
\label{tie-detection-prob}
\end{equation}
where $L$ is the arms'-length difference. 
This interference pattern is nonsinusoidal if $|k_1|\neq |k_2|$. It may be conveniently characterized by
its slope or \emph{sensitivity} with respect to small changes of the phase between the MZI arms, defined as \cite{NJP07}
\begin{equation}
S \equiv \frac{2}{k_{\rm max}}\left|\frac{d P_+}{ dL}\right|, \label{e3}
\end{equation} where $k_{\rm max}$ is the \emph{larger} of $k_1$ and $k_2$.

An unusual feature of this scheme is the length dependence of the superposition in (\ref{psi-tie}), which makes the \emph{internal state} change
with $L$. This fact may be used for obtaining which-way information: the  distinguishability of the two ways\cite{Vaidman,Englert},
as read from the internal state, is then found to be
\begin{equation}
D=2\left|c_1c_2\sin{\frac{(k_2-k_1)L}{2}}\right|.
\label{D-tie}
\end{equation}
Equations (\ref{tie-detection-prob})-(\ref{D-tie}) imply an unconventional tradeoff 
between two usually complementary tasks, namely, our \emph{joint} knowledge of the path  
and the phase shift between the MZI arms\cite{NJP07,IJPB06}. This may clearly be seen, e.g., 
for the case $k_2=3k_1$: counter-intuitively, near the point $k_1L/\pi=0.5$ the distinguishability
is $D\approx 1$ while the slope (sensitivity) $S$ of the interference fringe pattern at the \emph{same} phase value is far from
the expected $0$, this yielding wave-like information. 

We have proposed\cite{NJP07,IJPB06,PRL06} a possible realization of this TIE interferometry scheme using cold atoms in atomic 
waveguides. Although this realization is within current experimental possibilities, it still is quite challenging.

In this Letter we propose a simpler analog to TIE interferometry for a photon. 
The photon polarization defines its internal state, which determines, for well-defined 
orbital angular momentum, the spatial phase accumulated 
in the MZI.

\begin{figure}[htb]
%\centering \hspace{-0.06\linewidth}
%\includegraphics[width=.9\linewidth]{f-interferometer}
%\subfigure[]{
\includegraphics[width=0.95\linewidth]{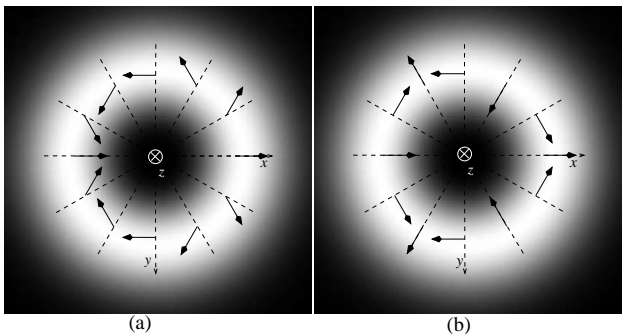}
%\subfigure[]{
%\includegraphics[width=0.6\linewidth]{F1}}
%\subfigure[]{
%\includegraphics[width=0.6\linewidth]{fig1b}}
\caption{ An example of the rotational symmetry of the Laguerre-Gaussian beam $LG_p^l$. 
Vector plot of the transverse beam profile for $p=0,\;l=2$. The arrows represent the electric
field vector at a given time and fixed propagation length for 
(a) right-hand polarized photon ($s=-1$, 1-fold symmetry), (b)
left-hand polarized photon ($s=+1$, 3-fold symmetry). } \label{beam-profile}
\end{figure}

Let us assume a circularly polarized photon with spin angular momentum 
$S=s\hbar$ ($s=\pm 1$ for $L$, $R$ polarization, respectively). This photon may be endowed with 
well defined orbital angular momentum (OAM) $L=l\hbar$, if the light beam is in either a 
Laguerre-Gaussian\cite{Allen92} $LG_p^l$ or a pseudo-nondiffracting Bessel-Gaussian mode\cite{Bouchal} $BG^l$. 
The field amplitude of the Laguerre-Gaussian 
mode $LG_p^l(r,\phi,z)$ or Bessel-Gaussian mode $BG^l(r,\phi,z)$ is proportional to
$\exp{(il\phi)}$, where $\phi$ is the angular coordinate in the transverse plane of the beam. 
This exponential factor is responsible for the well-defined
OAM\cite{Allen92,Bouchal,Dholakia-opt-com98}. 
The total angular momentum per photon\cite{Courtial98} is then $J=L+S$. This fact guarantees $(l+s)$-fold rotational 
symmetry of the transverse electric field (see fig.~\ref{beam-profile}). 
In other words, upon rotation of the beam through 
an angle $\alpha$, the phase shift of the photon is $\exp{[i(l+s)\alpha]}$. 
%%%%%%%%%%%%%%%%%%%%%% F I G U R E %%%%%%%%%%%%%%%%%%%%%%%%%%%%
%%%%%%%%%%%%%%%%%%%%%% F I G U R E %%%%%%%%%%%%%%%%%%%%%%%%%%%%
This is the key to our ability to acquire l-s correlated phase shifts
$\exp{[i(l+1)\alpha]}$, $\exp{[i(l-1)\alpha]}$ with $s=\pm 1$, respectively. 
%Hence we can establish
%the following analogy with Eq.~(\ref{psi-tie}): $k_2(k_1)\leftrightarrow l+s\;(l-s)$, $\ket{2}(\ket{1})\leftrightarrow
%\ket{R}(\ket{L})$. 

\begin{figure}[htb]
%\centering \hspace{-0.06\linewidth}
%\includegraphics[width=.9\linewidth]{f-interferometer}
%\subfigure[]{
\includegraphics[width=0.8\linewidth]{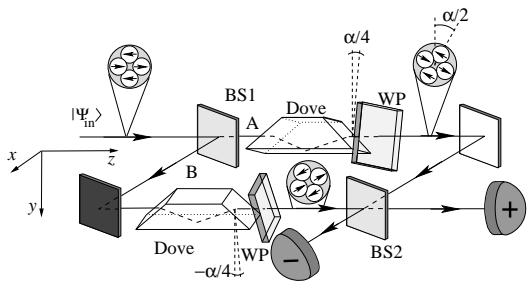}
%\subfigure[]{
%\includegraphics[width=0.6\linewidth]{F1}}
%\subfigure[]{
%\includegraphics[width=0.6\linewidth]{fig1b}}
\caption{ Schematic plot of the proposed realization. A particular example of state (\ref{psi-input}), $\ket{l}\otimes\ket{x}$, is injected into the MZI, 
formed by two 50\%-50\% (nonpolarizing) beam splitters BS1 and BS2 and two mirrors.  
After passing the rotated Dove prism and wave-plate (WP), the linear polarization is rotated 
by a different angle in each arm, depending on the interferometric phase. This allows us to infer 
which-way information by measuring the resulting photon polarization in basis (\ref{meas-basis}) by
detectors ($+$) and 
($-$) after BS2. } \label{MZI}
\end{figure}

As in previous experiments\cite{Allen92,Courtial98} we propose to use two optical components to 
manipulate the beam, namely, a wave plate (WP) to mirror the photon's linear
polarization and a Dove prism to flip the spatial profile of the beam with respect to the prism's bottom face plane. 
To describe the action of these respective components on the beam, we use the notation $LG_p^l\equiv \ket{l}$,
and define the right-hand oriented basis as follows (see fig.~\ref{MZI}): $x$ is parallel to the bottom
phase plane of the unrotated prism, $y$ is the outward normal to this plane, and $z$ is parallel to the
beam propagation direction, i.e. to the prism's longitudinal axis. When the Dove prism is
rotated around the propagation direction through an angle $\alpha$ with respect to the $x$ axis, its
action on the state $\ket{l}$ is 
\begin{equation}
\op{D}(\alpha)\ket{l}=\exp{(i2l\alpha)}\ket{-l}.
\label{dove-action}
\end{equation}
This can be deduced from the fact that a flip along the axis rotated through $\alpha$ causes
the transformation $\phi\rightarrow 2\alpha-\phi$.
\\During the passage of the photon through the prism, two refractions 
and a total reflection from the bottom face plane take place, affecting the
polarization state of the photon, according to Fresnel's
formula\cite{Born}. Here we represent this action by the operator $\op{P}_{\rm D}={\rm
diag}(d_x,d_y)$, in the basis of linear-polarization states $|x\rangle,~|y\rangle$ along $x$ and $y$.
The wave plate (WP) does not affect the beam profile, but does affect its polarization, via the operator
$\op{P}_{\rm W}={\rm diag}(w_x,w_y)$. The complex factors in $\op{P}_W$ are unimodular, i.e., $|w_x|=|w_y|=1$. 

Thus the joint action of the {\em unrotated} ($\alpha=0$) Dove prism and wave-plate on the input state is 
described by
\begin{equation}
\op{O_0} \equiv \op{D}(0)\otimes\op{P}_{\rm W}\op{P}_{\rm D}.
\label{joint-action}
\end{equation}
The operator (\ref{joint-action}) is defined on the space spanned by states $\ket{l}\otimes\ket{\psi}$, where
$\ket{\psi}$ is an arbitrary polarization state. 

After rotation of the Dove prism and the 
wave-plate around the $z$ axis through an angle $\alpha$, the operator (\ref{joint-action}) is transformed (rotated) to 
\begin{eqnarray}
\nonumber
\op{O_\alpha} \equiv \op{D}(\alpha)\otimes \op{U}\op{P}_{\rm W}\op{P}_{\rm
D}\op{U}^\dagger= \label{rotated-joint-action} \nonumber \\
\op{D}(\alpha)\otimes\left [a\op{I}+b\op{P}_{\rm HW}(\alpha)\right].
\label{rotated-joint-action-simple}
\end{eqnarray}
Here, $\op{U}=\exp{(-i\op{\sigma}_y\alpha)}$, $\pmb I$ is the identity operator, 
$\op{P}_{\rm HW}(\alpha)=\op{U}\op{P}_{\rm HW}\op{U}^\dagger$ is a   
half-wave plate rotation (HWP), with $\op{P}_{\rm HW}={\rm diag}(1,-1)$; 
$a$, $b$ are functions of $d_x$, $d_y$, $w_x$, $w_x$ normalized by $|a|^2+|b|^2=1$. They can be made such
that $|b/a|\gtrsim 20$. This ratio can be enlarged still further using antireflective coating of the
prism. We can then eliminate the unwarranted influence of the $a\op{I}$ term in 
(\ref{rotated-joint-action-simple}), safely neglecting $|a|\ll 1$ compared to $|b|\simeq 1$. We then 
describe the action of (\ref{rotated-joint-action-simple}) on the states 
$\ket{l}\otimes\ket{R(L)}$, where the right (left) polarized states are $\ket{R(L)}=(\ket{x}\mp i\ket{y})/\sqrt{2}$, as
\begin{eqnarray}
\nonumber
\op{D}(\alpha)\otimes\op{P}_{\rm
HW}(\alpha)[\ket{l}\otimes\ket{R}]=
\exp{[i2(l-1)\alpha]}\ket{-l}\otimes\ket{L}],\\
\op{D}(\alpha)\otimes\op{P}_{\rm
HW}(\alpha)[\ket{l}\otimes\ket{L}]=
\exp{[i2(l+1)\alpha]}\ket{-l}\otimes\ket{R}].
\label{e8}
\end{eqnarray}
One can see that after the passage through {\em both} rotated optical elements, the states
$\ket{L}$, $\ket{R}$ acquire phase factors $\exp{[i2(l\pm1)\alpha]}$, respectively. 
We can then establish the correspondence with Eq.~(\ref{psi-tie}) for TIE states as follows
\begin{eqnarray}
\nonumber
\ket{1}&\leftrightarrow&\ket{l}\otimes\ket{R},\;\ket{2}\leftrightarrow\ket{l}\otimes\ket{L},\\
k_1L&\leftrightarrow&(l-1)\alpha,\;k_2L\leftrightarrow(l+1)\alpha.
\label{analogy}
\end{eqnarray}

We may now proceed towards our goal: interferometry with polarized photons having well defined OAM.
Let us use 
\begin{equation}
\ket{\Psi_{\rm in}}=\ket{l}\otimes\ket{\psi}=\ket{l}\otimes(c_1\ket{R}+c_2\ket{L})
\label{psi-input}
\end{equation}
as an input state of the MZI (see fig.~\ref{MZI}), where $\ket{\psi}$ is a well-defined
polarization state of the photon. 
%%%%%%%%%%%%%%%%%%%%%% F I G U R E %%%%%%%%%%%%%%%%%%%%%%%%%%%%
%%%%%%%%%%%%%%%%%%%%%% F I G U R E %%%%%%%%%%%%%%%%%%%%%%%%%%%%
Assuming the MZI arms length
difference to be zero, the only factors determining the interferometric phase are those of (\ref{e8}) due to rotations of
the Dove prism and wave-plate around their joint axis (see fig.~\ref{MZI}). After the state
(\ref{psi-input}) passes BS1, we apply the transformations (\ref{e8}) to each arm, i.e. rotate the
elements through angle $\alpha/4$ in arm $A$ and $-\alpha/4$ in arm $B$. 

The polarization states in each arm just before BS2 are then 
\begin{eqnarray}
\nonumber
\ket{\psi_A}&=&\op{P}_{\rm HW}(\alpha/4)\ket{\psi},\\
\ket{\psi_B}&=&\op{P}_{\rm HW}(-\alpha/4)\ket{\psi},
\label{final-states}
\end{eqnarray}
To obtain which-way information from the MZI output, the optimal measurement of the photon polarization states 
is found to be in the basis rotated by $\pm 45^\circ$ relative to $x$ only
\begin{equation}
\ket{\pm45}=\left(\ket{x} \pm \ket{y}\right)/\sqrt{2}.
\label{meas-basis}
\end{equation}
This can be done by the use of polarizing beam splitters in front of the output detectors (+) and (--). Then, one of
{\em four} detection events may occur for each photon, corresponding to $\ket{+}\ket{+45}$, $\ket{-}\ket{+45}$, $\ket{+}\ket{-45}$, and 
$\ket{-}\ket{-45}$ where $|\pm\rangle$ denote the states localized at ports ($+$) or ($-$), respectively.
Regardless whether $\ket{+}$ or $\ket{-}$ is detected, when obtaining $\ket{+45}$, our guess is that
the photon took arm A, and when $\ket{-45}$, that the photon took arm B.

The \emph{total} detection probability (irrespective of polarization) at the $(+)$ output port oscillates with $\alpha$ as 
%\begin{subequations}
\begin{eqnarray}
\nonumber
P_+&=&\frac{1}{2}+\frac{|c_1|^2\cos{[(l-1)\alpha]}+|c_2|^2\cos{[(l+1)\alpha]}}{2}.\\
\label{plus-prob-new}
\end{eqnarray}
The corresponding sensitivity (slope) is given by
\begin{eqnarray}
S&=&\frac{2}{(l+1)}\left|\frac{dP_+}{d\alpha}\right|\\
\nonumber
&=&\left||c_1|^2\frac{(l-1)}{(l+1)}\sin{[(l-1)\alpha]}+|c_2|^2\sin{[(l+1)\alpha]}\right|. 
\label{e14}
\end{eqnarray}
%\end{subequations}
We may calculate the distinguishability in the same way as in Eq.~(\ref{D-tie}), to be
\begin{equation}
D=2\left|c_1c_2\sin{(\alpha)}\right|,\;{\rm for\;all}\;l.
\label{D-tie-new}
\end{equation}
This distinguishability may be related, via the relation\cite{Wooters,Vaidman,Englert,Rempe,Kwiat,NJP07} $D=2\mathcal{L}-1$, 
to the likelihood of a correct which-way guess
\begin{eqnarray}
\mathcal{L}=|\langle+45\ket{\psi_A}|^2=|\langle-45\ket{\psi_B}|^2.
\label{likelihood}
\end{eqnarray}

The analogy of Eqs.~(\ref{tie-detection-prob}), ({\ref{e3}) and  (\ref{plus-prob-new}), (\ref{e14}), on the one hand, and Eqs.~(\ref{D-tie}) and 
(\ref{D-tie-new}) on the other, implies that we obtain for polarized photons with OAM the same unusual effects previously predicted for 
TIE states of massive particles\cite{NJP07}.
Using beams with different OAM (topological charge) number $l$, we can effectively vary the 
degree of ``violation'' of the standard complementarity as for 
TIE states with wavenumber ratios\cite{NJP07} $k_2/k_1=\kappa$. Namely, $\kappa=[-1,2,3,\infty]$ correspond to $l=[0,3,2,1]$ (fig.~\ref{detection-prob}).

%%%%%%%%%%%%%%%%%%%%%% F I G U R E %%%%%%%%%%%%%%%%%%%%%%%%%%%%
\begin{figure}[htb]
\centering \hspace{-0.05\linewidth}
\subfigure[]{
\includegraphics[width=0.5\linewidth]{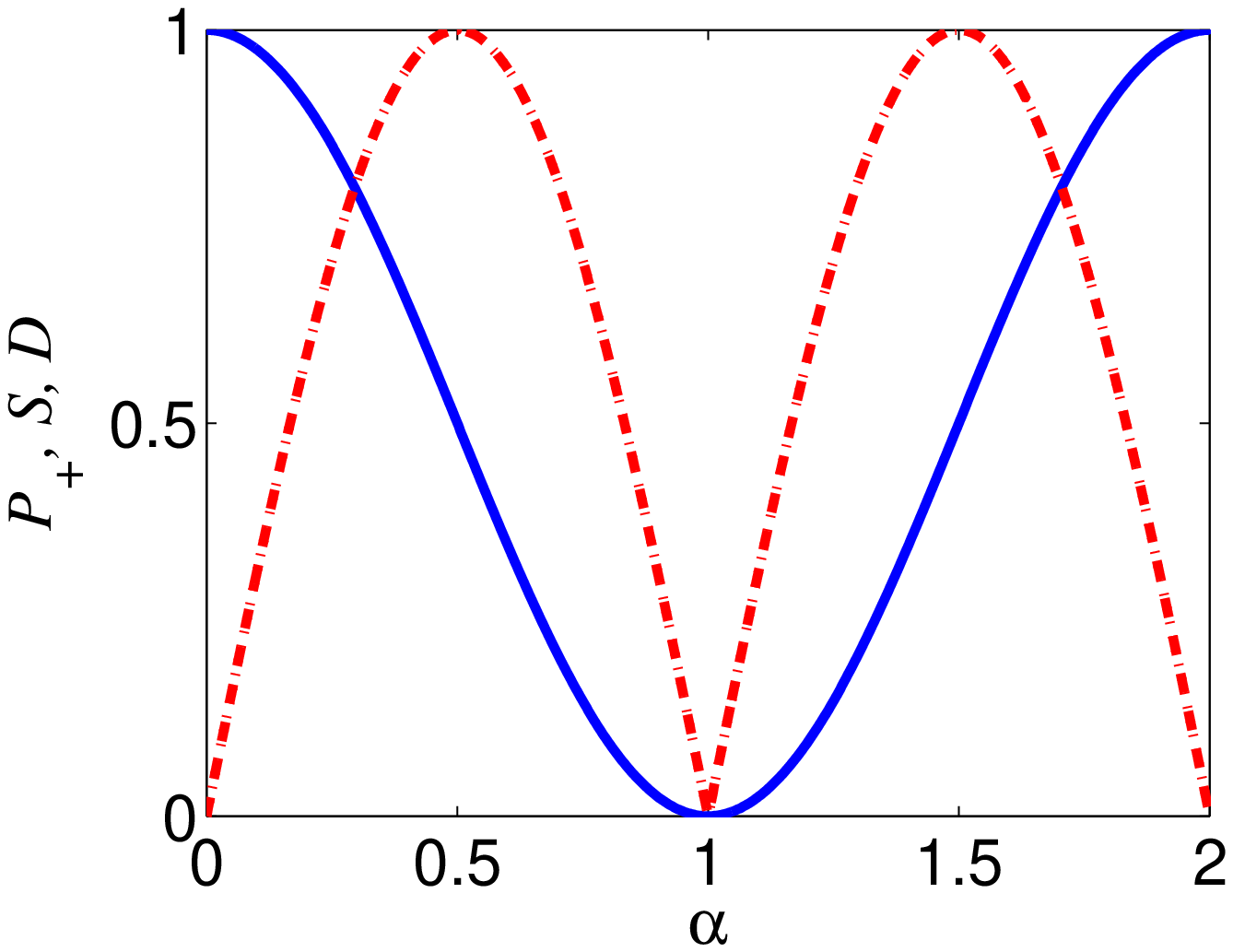}}
\subfigure[]{
\includegraphics[width=0.5\linewidth]{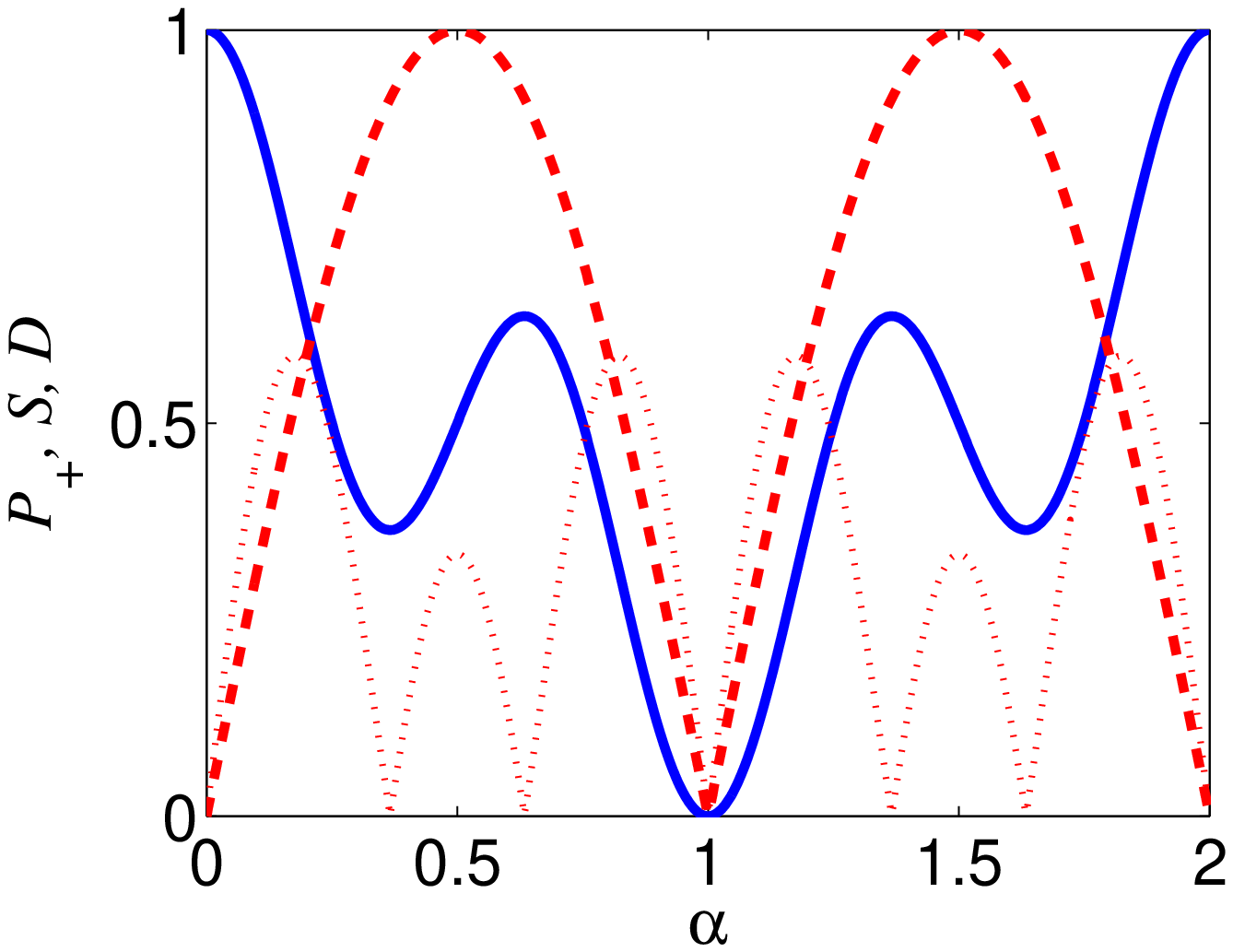}}
%\subfigure[]{
%\includegraphics[width=0.6\linewidth]{fig1b}}
\caption{ Detection probability $P_+$ (solid), phase sensitivity $S$ (dotted)
and distinguishability $D$ (dashed) as a function of $\alpha$ for $c_1=c_2=1/\sqrt{2}$, 
for: (a) $l=0$ ($S$ and $D$ are matching each other) and (b) $l=2$ (Eqs. (\ref{plus-prob-new}-\ref{D-tie-new})).}
\label{detection-prob}
\end{figure}
%%%%%%%%%%%%%%%%%%%%%% F I G U R E %%%%%%%%%%%%%%%%%%%%%%%%%%%%

In conclusion, we have proposed a setup which uses a Dove prism and a wave plate to
manipulate the polarization state of a photon in a beam with well defined
orbital angular momentum (OAM). The rotation of both optical components causes the two orthogonal
{\em circular} polarizations to acquire spatial phases with the ratio $(l+1)/(l-1)$, $l$ being the \emph{topological
charge} of the beam. Using such a polarized photon with OAM as an input to a Mach-Zehnder
interferometer (MZI) with arms of nearly equal length, we can concurrently guess 
which way the photon took in the MZI and which was the sign of a small phase shift between the MZI arms.
Using our setup gives results far superior to those following from the standard complementarity
relation\cite{Vaidman,Englert}. For example, to detect a phase shift $(l+1)\alpha\approx 10^{-2}$ requires
$9\cdot 10^{4}$ photons on average, for $l=2$, among which less than {\em one} photon on average will have
the path determined incorrectly. Within the bounds of standard
complementarity, we would need $5.3\cdot 10^{4}$ photons, out of which about $2.6\cdot 10^{3}$ would have
the path determined incorrectly\cite{NJP07}.

We thank G. Rempe, A. Zeilinger, M. $\dot{\rm Z}$ukowski, R. \v Celechovsk\' y, A.V. Zhukov, and J. K\v repelka for useful discussions and 
acknowledge the support of GA\v CR (GA202/05/0486), 
M\v SMT (LC 06007), MSM~(6198959213), GIF, ISF and EC (SCALA).

%%%%%%%%%%%%%%%%%%%%%% F I G U R E %%%%%%%%%%%%%%%%%%%%%%%%%%%%
%\begin{figure}[h]
%\centering
%\includegraphics[width=0.6\linewidth]{F2}
%\caption{The ${\cal S}$-${\cal D}$ dependence at $\phi_m$ (Eq.) for standard case $\kappa=1$, Eq.() (dashed-blue) and the PPE cases
%Eq.(\ref{ellipse}) 
%$\kappa=3$ (red) and $\kappa\gg 1$ Eq.() (green).}
%\label{f2}
%\end{figure}
%%%%%%%%%%%%%%%%%%%%%% F I G U R E %%%%%%%%%%%%%%%%%%%%%%%%%%%%

%\hrulefill

\end{document}